\documentclass[pr1,twocolumn,longbibliography]{revtex4-1}
\usepackage{amsmath,amssymb,latexsym,epsfig,graphics,epsf}
\usepackage{graphics}
\usepackage{float}

\newcommand{\bR}{\bold R}
\newcommand{\bX}{\bold X}
\newcommand{\bY}{\bold Y}
\newcommand{\bRa}{{\bf R}_{\rm a}}
\newcommand{\bXa}{{\bf X}_{\rm a}}
\newcommand{\bYa}{{\bf Y}_{\rm a}}
\newcommand{\bRb}{{\bf R}_{\rm b}}
\newcommand{\bXb}{{\bf X}_{\rm b}}
\newcommand{\bYb}{{\bf Y}_{\rm b}}

\newcommand{\be}{\begin{equation}}
\newcommand{\ee}{\end{equation}}

\newcommand{\Sex}{{S}_{\rm ex}}

\begin{document}

\title{``Pseudoisomorphs'' in liquids with intramolecular degrees of freedom}
\date{\today}
\author{Andreas Elmerdahl Olsen}
\author{Jeppe C. Dyre}
\author{Thomas B. Schr{\o}der}\email{tbs@ruc.dk}
\affiliation{Glass and Time, IMFUFA, Department of Science and Environment, Roskilde University, P.O. Box 260, DK-4000 Roskilde, Denmark}

\begin{abstract}
Computer simulations show that liquids of molecules with harmonic intramolecular bonds may have ``pseudoisomorphic'' lines of approximately invariant dynamics in the thermodynamic phase diagram. We demonstrate that these lines can be identified by requiring scale invariance of the inherent-structure reduced-unit low-frequency vibrational spectrum evaluated for a single equilibrium configuration. This rationalizes why excess-entropy scaling, density scaling, and isochronal superposition apply for many liquids with internal degrees of freedom.
\end{abstract}

\maketitle

Some condensed matter systems are simpler than others. Thus metallic liquids and solids, as well as van der Waals bonded and weakly ionic or dipolar organic systems, generally have simple structure and dynamics of their atoms or molecules \cite{barrat,kir07,rol10}. These systems also obey a number of empirical freezing and melting rules \cite{lin10,sti75,cha07a,han70}. In contrast, systems with strong directional bonds like covalently or hydrogen-bonded systems are often quite complex, for instance by having many different crystal structures, by having in parts of their thermodynamic phase diagram a diffusion constant that increases upon isothermal compression, by melting instead of freezing upon compression, etc \cite{cha09}. Water is exceedingly complex and disobeys many rules that apply for metals and van der Waals systems \cite{deb03}. 

Simple atomic systems studied for many years by computer simulations include the Lennard-Jones (LJ) pair-potential system, the hard-sphere system, the Yukawa (screened Coulomb) system, inverse power-law systems, etc \cite{ros77,ros00,hey08,pon11a,han13}. Several molecular models are also simple in the sense that they have lines in the thermodynamic phase diagram, so-called isomorphs, along which the dynamics is invariant to a good approximation in suitably reduced units \cite{IV,ing12b,dyr14}. This implies that the phase diagram becomes effectively one-dimensional in regard to many properties which prohibits anomalous behavior, and also that these systems follow certain scaling laws \cite{IV,ing12,dyr14,dyr16}. Systems with isomorphs are sometimes referred to as R (Roskilde) simple \cite{mal13,abr14,fer14,fle14,pra14,sch14,buc15,har15,hey15,sch15} to distinguish this class from systems defined by additive pair potentials \cite{han13}.

The first scaling law was reported by Rosenfeld, who back in the 1970s discovered excess-entropy scaling \cite{ros77}. Recall that the excess entropy, $\Sex$, is defined as the entropy minus that of an ideal gas at the same density and temperature, a quantity that is negative because any system is less disordered than the ideal gas. From state-of-the art computer simulations of his time Rosenfeld demonstrated that along the lines of constant $\Sex$ transport coefficients are invariant for a number of atomic models, including the LJ system. Judging from the Science Citation Index this discovery was not appreciated by the broad scientific community until about fifteen years ago. Since then excess-entropy scaling has been reported in many different contexts \cite{san01,dal06,mit06,goe08,kre09,aga10,cho10c,aga11,bar11,gal11,pon11,hel12,voy13,abr14,ban16,cao16,jak16}. Rosenfeld justified excess-entropy scaling by assuming the system in question can be approximated by a state-point dependent hard-sphere system, an argument that is difficult to generalize to elongated or flexible molecules.

A scaling law that was first observed in experiments is the so-called density scaling according to which the average relaxation time of a supercooled van der Waals bonded liquid to a good approximation is invariant along the lines of constant $\rho^\gamma/T$, where $\rho$ is the  density, $T$ is the temperature, and $\gamma$ is an empirical exponent \cite{alb04,cas04a,tar04,rol05,rol10}.  A further and highly nontrivial characteristic of van der Waals liquids is isochronal superposition, i.e., the observation that the average relaxation time determines the entire relaxation time spectrum \cite{nga05,nie08}.

For R simple systems excess-entropy scaling, density scaling, and isochronal superposition are explained by the isomorph theory \cite{IV, dyr14}, which predicts the existence of lines of invariant dynamics in the phase diagram whenever the potential-energy function $U(\bR)$ obeys the scaling condition $U(\bRa)<U(\bRb) \Rightarrow U(\lambda\bRa)<U(\lambda\bRb)$ ($\bR$ is the vector of all particle coordinates and $\lambda$ is a scaling parameter) \cite{IV,sch14}. This condition is trivially and rigorously obeyed for any Euler-homogeneous function plus a constant, but less trivially so also to a good approximation for the above-mentioned simple atomic system as well as for R simple models of rigid molecules \cite{sch14}. 

Henceforth, reduced units refer to the thermodynamic state-point dependent units based on the length $\rho^{-1/3}$, the energy $k_BT$, and the time $\sqrt{m\rho^{-2/3}/k_BT}$ in which $m$ is the particle mass. The isomorph theory predicts invariant dynamics in reduced units along  isomorphs defined as lines of constant $\Sex$ \cite{sch14}, implying that R simple systems obey excess-entropy scaling as well as isochronal superposition. Density scaling is an approximation to a more general scaling law predicted by the isomorph theory \cite{boh12,dyr14}.

A system has isomorphs if and only if it has strong correlations between its constant-density equilibrium virial and potential-energy fluctuations \cite{IV,sch14}, so only for liquids with strong correlations does the isomorph theory explain excess-entropy scaling, density scaling, and isochronal superposition. There are, however, results from both experiments \cite{tol01,rol03,alb04,nga05,rol05,rol10,abr14} and simulations \cite{kre09,aga10,cho10c,aga11,pon11,vel15a} showing that these and similar scaling laws apply to a larger class of liquids. Focusing for the moment on excess-entropy scaling, Truskett, Errington, and coworkers have proposed two different generalizations beyond simple atomic systems. By redefining the relevant reduced units it is possible to extend excess-entropy scaling to mixtures and soft ``atomic'' particles like the Gaussian core model \cite{kre09}. Another generalization, which is relevant for molecular models with chemical bonds represented by springs, is to define the excess entropy by subtracting the ideal-gas entropy of the flexible molecules in which way the spring contribution to the entropy is eliminated, see Ref. \onlinecite{cho10c} and references therein. This works well, but it is not clear why.

The importance of understanding the scaling properties of systems with flexible degrees of freedom is emphasized by the fact that many of the liquids found in experiment to obey density scaling and isochronal superposition are expected not to have strong virial potential-energy correlations due to bending and stretching intramolecular degrees of freedom. We explain below why liquids with internal degrees of freedom, \textit{in casu} harmonic bonds, can nevertheless have lines in the thermodynamic phase diagram of approximately invariant dynamics, and we demonstrate a method to identify these lines of invariant dynamics from a single equilibrium configuration.

\begin{figure}[htbp]
	\centering
	\includegraphics[width=.35\textwidth]{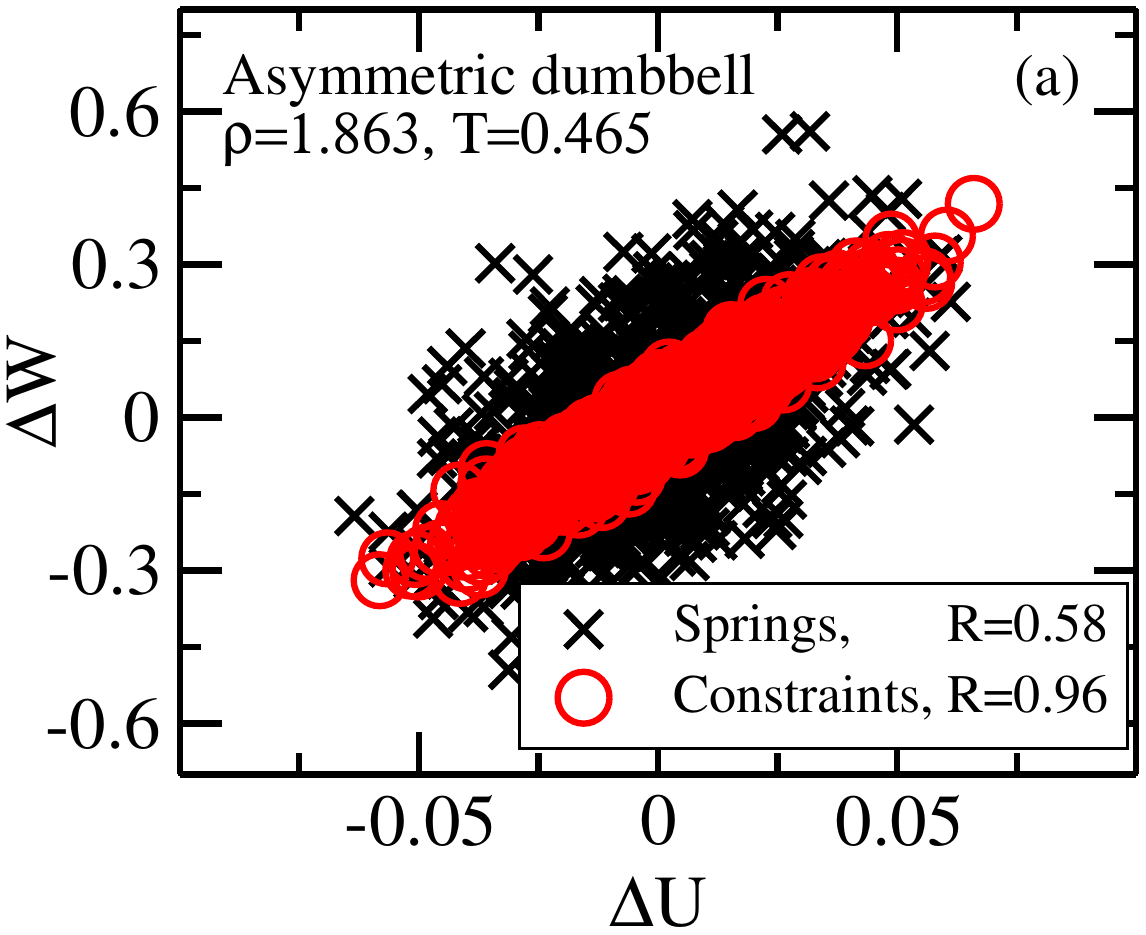}
	\includegraphics[width=.35\textwidth]{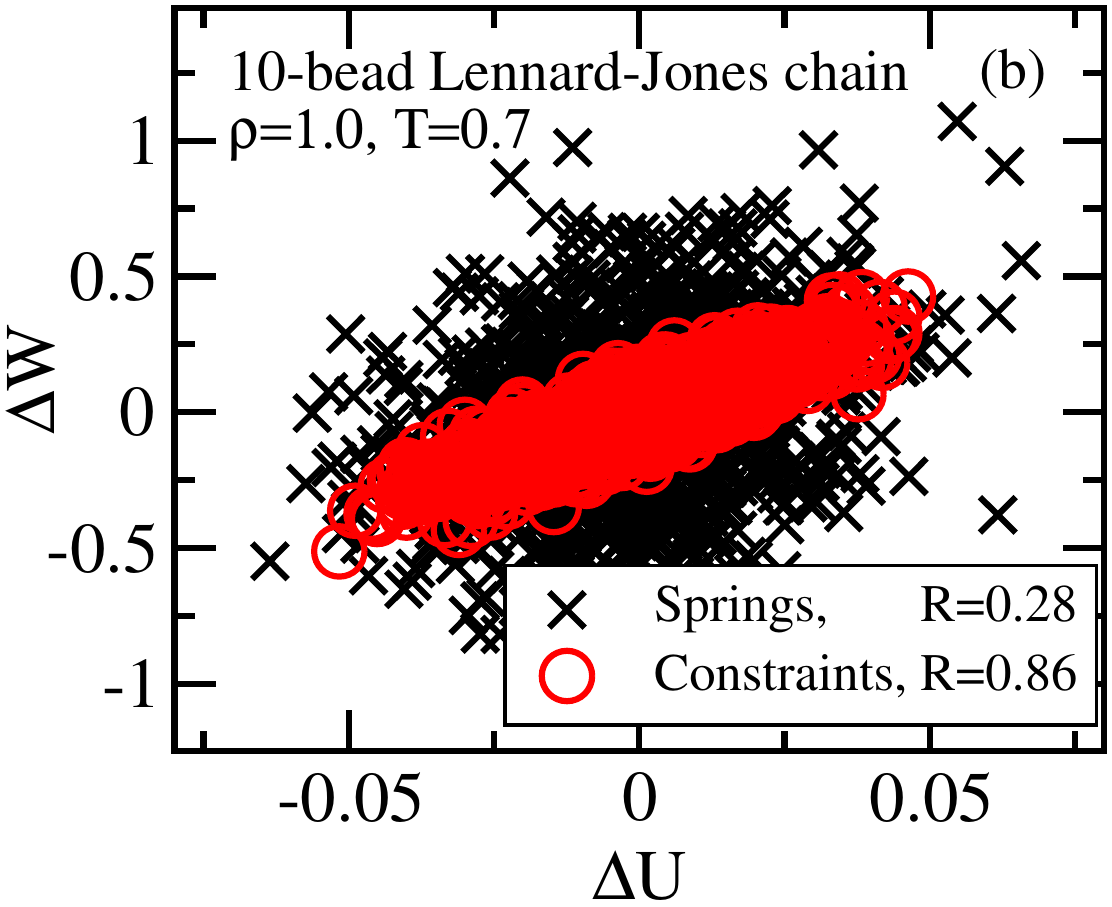}
	\caption{\label{fig:UWcorr}
Virial potential-energy scatter plots of the asymmetric dumbbell model (512 molecules) and the 10-bead Lennard-Jones chain model (200 chains) at the indicated thermodynamic state points (canonical ensemble data). The black crosses give results for the harmonic-bond case studied below, the red circles give results when the harmonic bonds are rigid, i.e., replaced by constant-length constraints. Only in the latter case do the models have strong virial potential-energy correlations \cite{sch09,vel14}.}
\end{figure}

We simulated two molecular models with harmonic bonds, the asymmetric dumbbell (ASD) model and the 10-bead Lennard-Jones chain (LJC) model. The ASD model consists of two different-sized LJ particles, the LJC model of a ten-bead chain of LJ particles; the simulated models are identical to their rigid-bond analogs discussed in Refs. \onlinecite{sch09,vel14} except that the bonds are harmonic springs (both spring constants are $3000$ in LJ units). The calculations were carried out using the RUMD GPU code \cite{rumd15} for standard Nose-Hoover NVT simulations.

Figure \ref{fig:UWcorr} shows scatter plots of virial versus potential energy in equilibrium at typical state points of the two models (black crosses). The Pearson correlation coefficients are, respectively, 0.58 and 0.28, which are far below the lower value $\sim 0.9$ characterizing systems with isomorphs \cite{IV}.

\begin{figure}[htbp]
\centering	
\includegraphics[width=.35\textwidth]{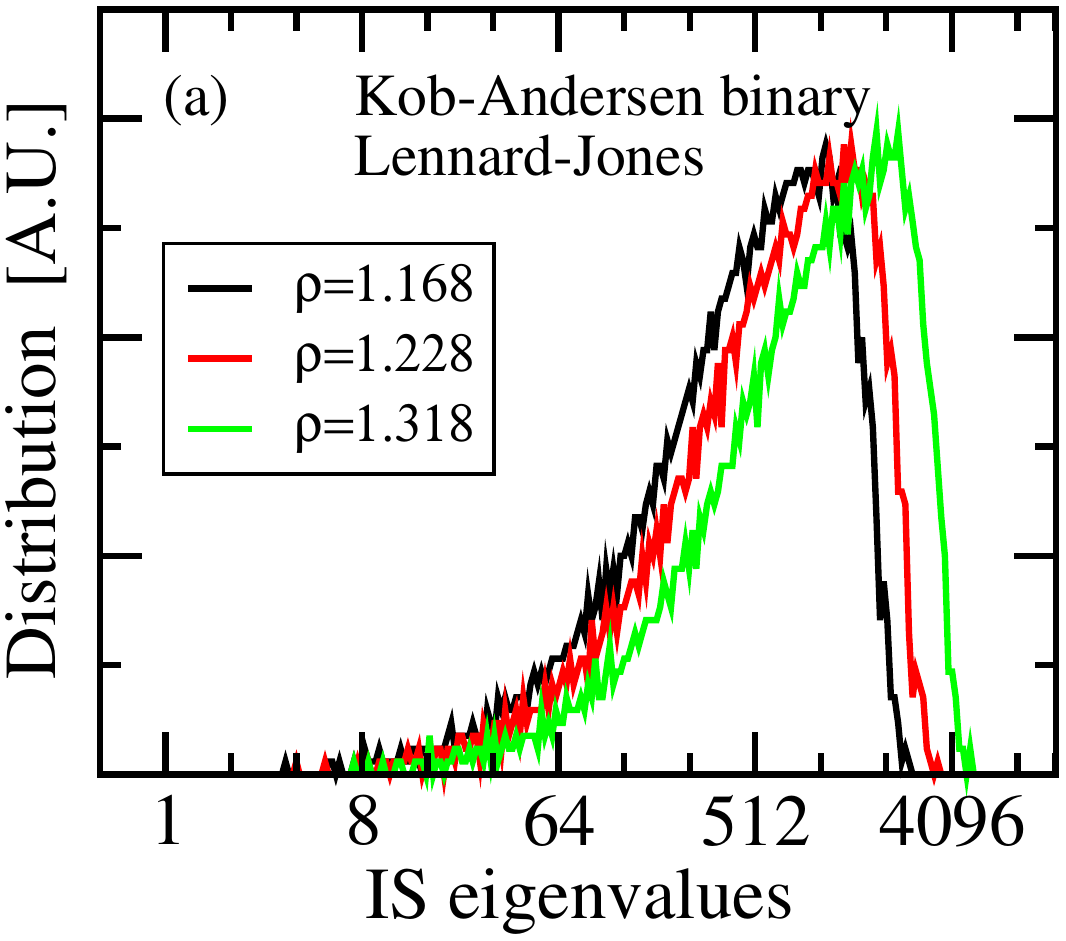}
\includegraphics[width=.35\textwidth]{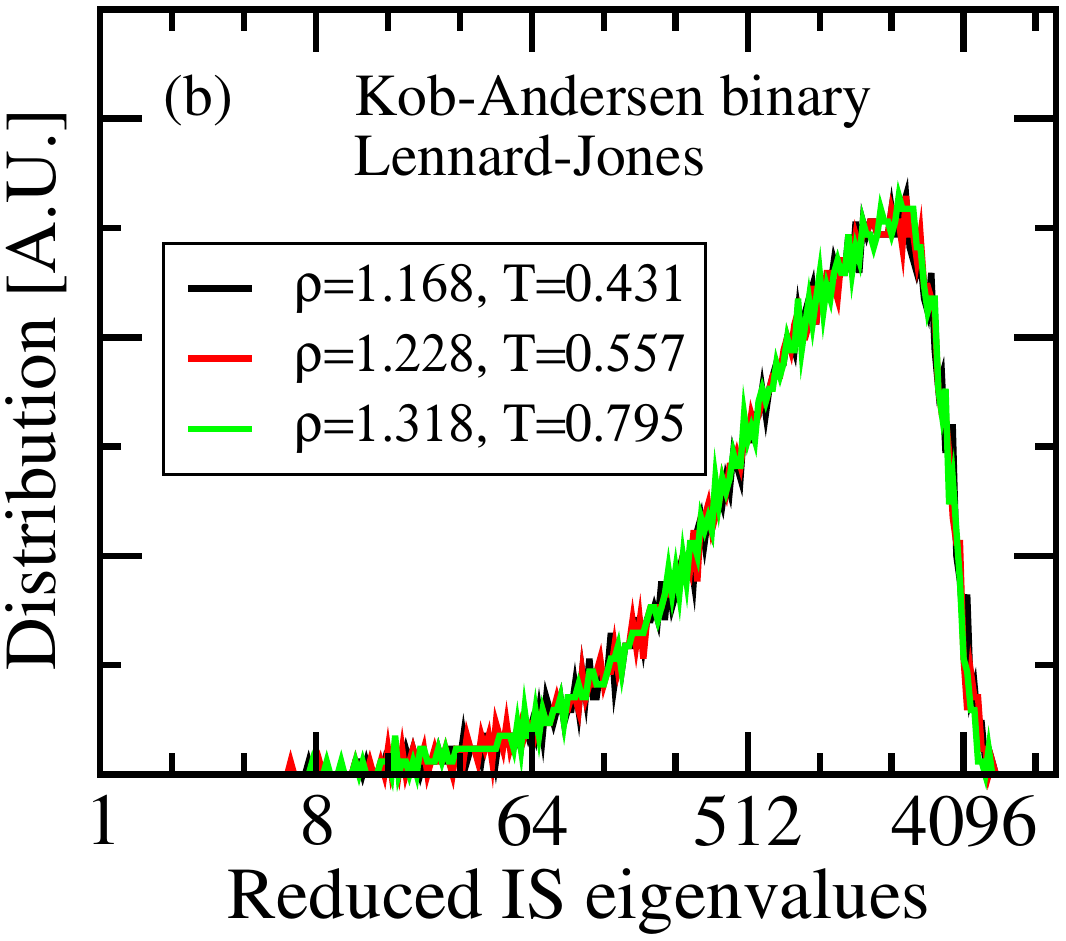}
\caption{\label{fig:SpectraKA}
Spectra of the inherent structure (IS) Hessian eigenvalues of the Kob-Andersen binary Lennard-Jones mixture (1000 particles), i.e., the distribution of eigenvalues of the symmetric 3000$\times$3000 matrix of second derivatives of the potential energy function at a potential-energy minimum.
(a) Red curve: Spectrum for one equilibrium configuration at the state point $(\rho,T)=(1.228,0.557)$ (LJ units) quenched to its inherent structure. Black and green curves: Spectra evaluated after uniformly scaling the configuration to the density in question, subsequently quenching it to the inherent structure.
(b) Same spectra in reduced units using the temperatures making the three state points isomorphic \cite{IV}. The three means of the logarithm of the reduced eigenvalues are given by, respectively, $|\ln{\tilde \lambda}|=6.75,6.74,6.73$.}
\end{figure}

To develop a practical method for identifying lines of invariant dynamics, henceforth referred to as ``pseudoisomorphs'' \cite{vel14}, we reason as follows. In their rigid bond versions the ASD and LJC models have strong virial potential-energy correlations (Fig. \ref{fig:UWcorr}, red symbols) and, consequently, isomorphs \cite{sch09,vel14}. This shows that the vibrational degrees of freedom are responsible for the breakdown of strong correlations. In order to eliminate the effect of these degrees of freedom on the scaling properties we take inspiration from the Kob-Andersen binary LJ  mixture, which has strong correlations and thus isomorphs \cite{IV}. Consider three isomorphic state points identified in Ref. \onlinecite{IV}. At one of these, $(\rho,T)=(1.228,0.557)$ in LJ units, we selected a configuration from an equilibrium $NVT$ simulation. The configuration was quenched to its inherent structure, i.e., local potential-energy minimum \cite{sti83}, at which the eigenvalues of the Hessian were identified. We then scaled the original configuration uniformly up and down to the two other densities, 1.168 and 1.318, and applied the same procedure. The three resulting eigenvalue spectra are shown in Fig. \ref{fig:SpectraKA}(a). Figure \ref{fig:SpectraKA}(b) shows that when the spectra are plotted in reduced units, they are virtually identical. This reflects the fact that for systems with isomorphs, to a good approximation the entire high-dimensional potential-energy surface undergoes a simple affine scaling upon a density change \cite{dyr14,sch14}. On the other hand, if not known in advance, the temperatures at the densities 1.168 and 1.318 making the three state points belong to the same  isomorph could have been identified by \textit{requiring} collapse of the reduced-unit spectra. This observation, incidentally, gives the first proof that isomorphs may be identified from a single configuration. -- We note that, in a related context, Gendelman, Pollack, and Procaccia recently showed that a snapshot of the particles of a zero-temperature amorphous solid is enough to determine the interparticle forces \cite{gen16}.

\begin{figure}[htbp]
\centering	
\includegraphics[width=.35\textwidth]{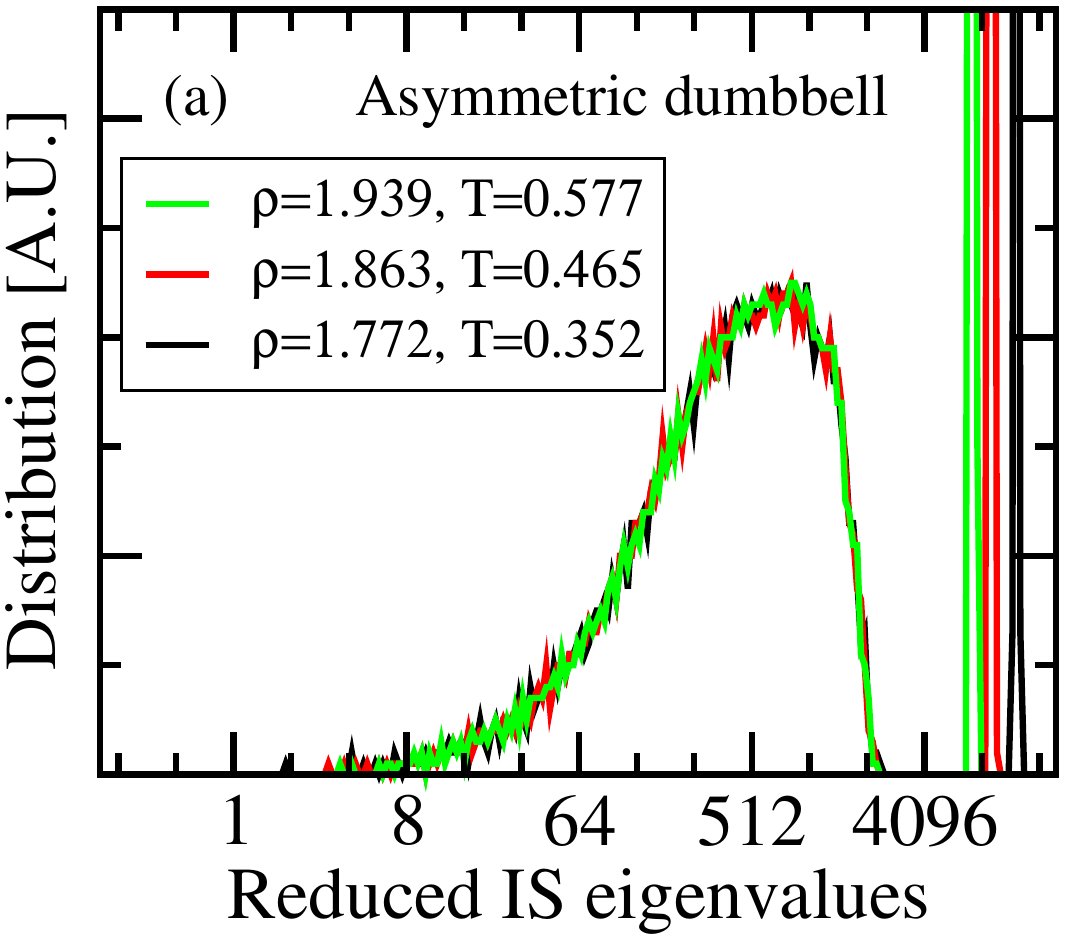}
\includegraphics[width=.35\textwidth]{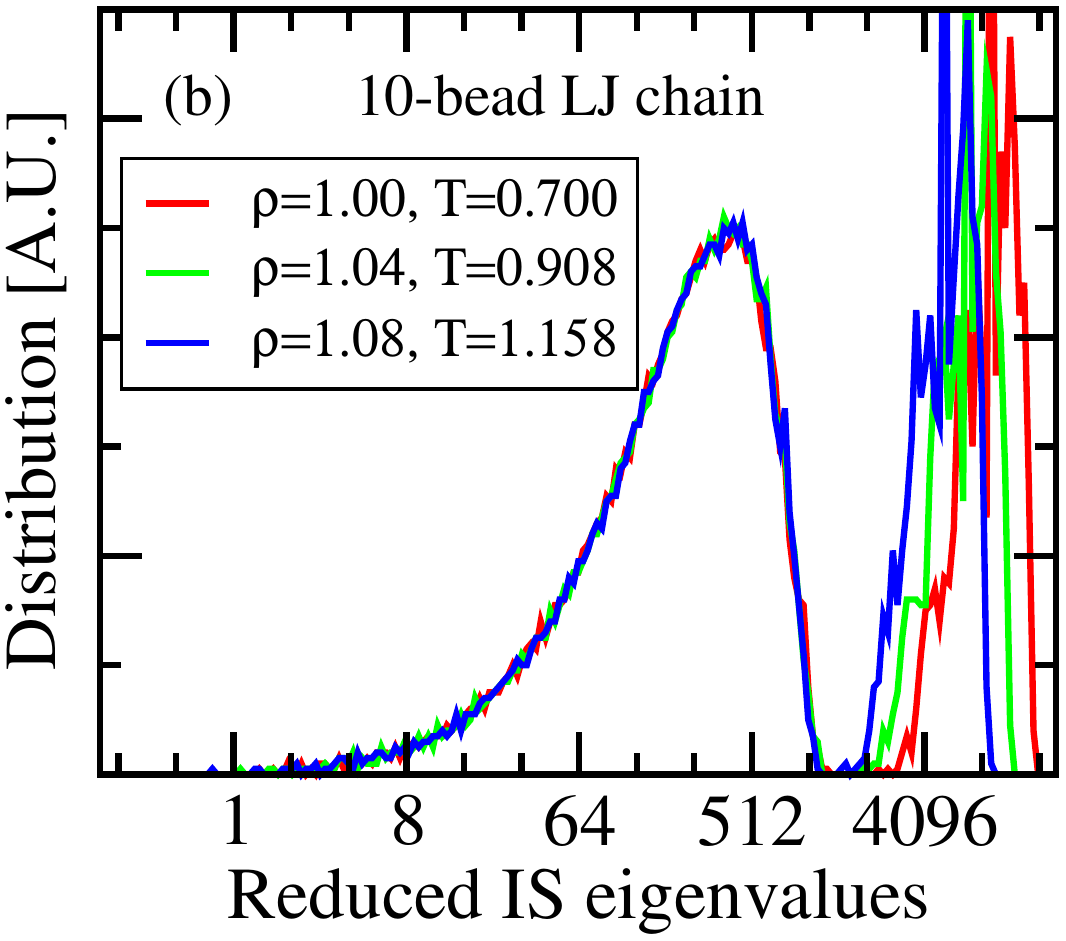}
\caption{\label{fig:Spectra}
(a) Reduced-unit vibrational spectra of the harmonic-bond asymmetric dumbbell model based on a configuration taken from an equilibrium simulation at the state point $(\rho,T)=(1.863,0.465)$. The ``pseudoisomorph'' temperatures at the two other densities were identified as described in the text by requiring invariance of the low-frequency part of the reduced spectra.
(b) Results when the same procedure is applied for the 10-bead Lennard-Jones chain with harmonic bonds, based on the reference state point $(\rho,T)=(1.0,0.7)$.
}
\end{figure}

The idea is now to apply the same procedure to the ASD and LJC harmonic-spring models with the expectation that the reduced eigenvalue spectra separate into a low-frequency ``scaling'' part and a high-frequency ``non-scaling'' part deriving from the bonds. At a given density the pseudoisomorphic temperature, i.e., the temperature that gives the same dynamics as that of a reference state point, is determined by requiring identical spectra in reduced units, excluding the high-frequency part of the spectra originating from the springs. To have a robust procedure, instead of visually fitting to collapse we calculate the temperature by requiring identical low-frequency vibrational entropy at the inherent structures in question. At the given density the pseudoisomorph temperature is thus found by requiring invariance of the sum of the logarithm of the reduced-unit Hessian eigenvalues, excluding the highest-frequency modes. Figures \ref{fig:Spectra}(a) and (b) show the resulting spectra in reduced units. There is excellent collapse of the entire low-frequency parts of the spectra. For both models the number of non-scaling, high-frequency eigenvalues is equal to the number of springs, confirming the physical interpretation of the origin of the high-frequency eigenmodes.

In principle, the method requires just a single equilibrium configuration provided, of course, the system is large enough. In practice, one will use several configurations to estimate the stability of the procedure. We did this for 256 independent configurations and found the relative standard deviations of the predicted temperatures to be $0.2\%$ (ASD) and $0.5\%$ (LJC).
	
	\begin{figure*}[htbp]
		\centering
		\includegraphics[width=.4\textwidth]{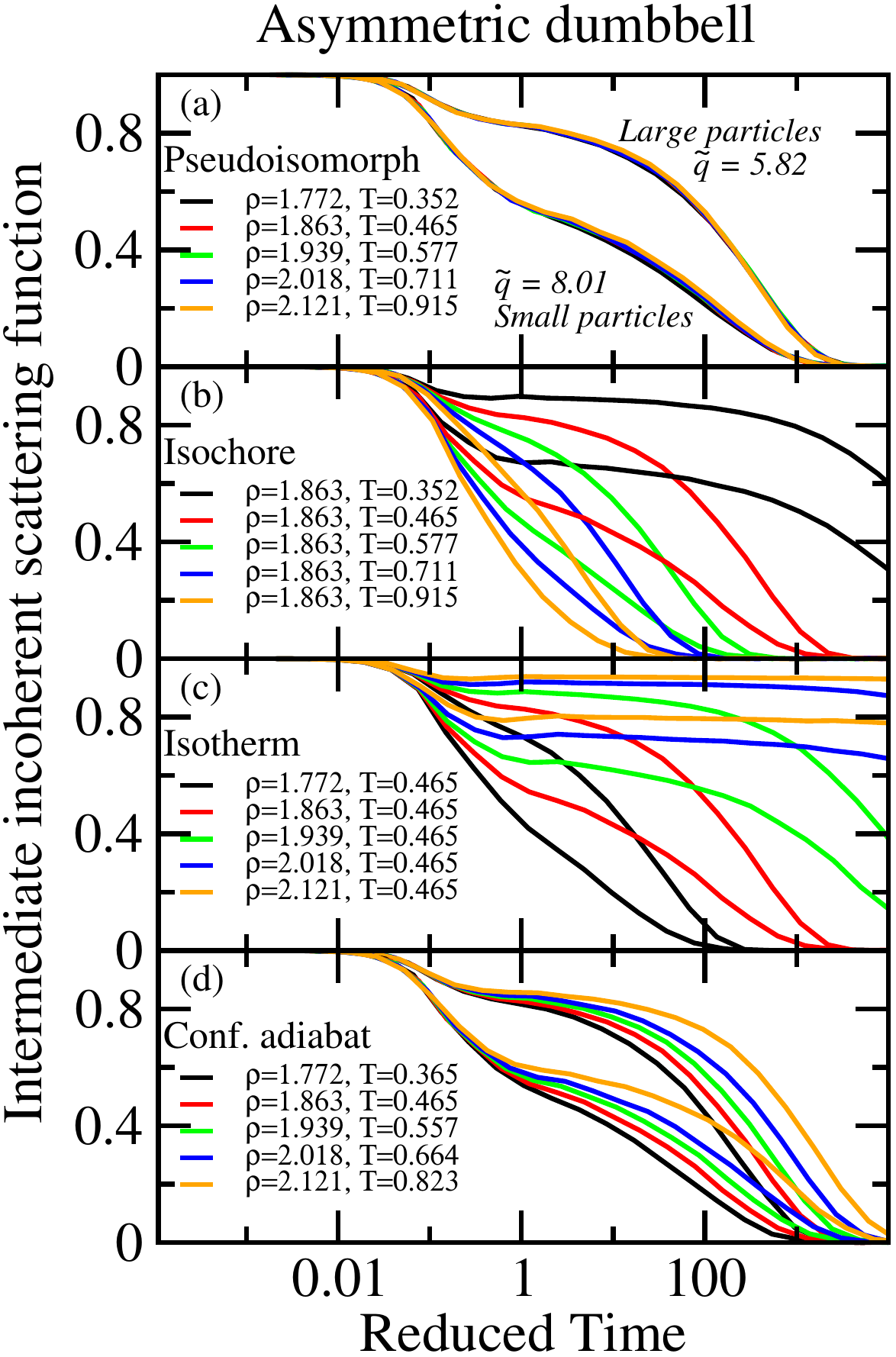}
		\includegraphics[width=.4\textwidth]{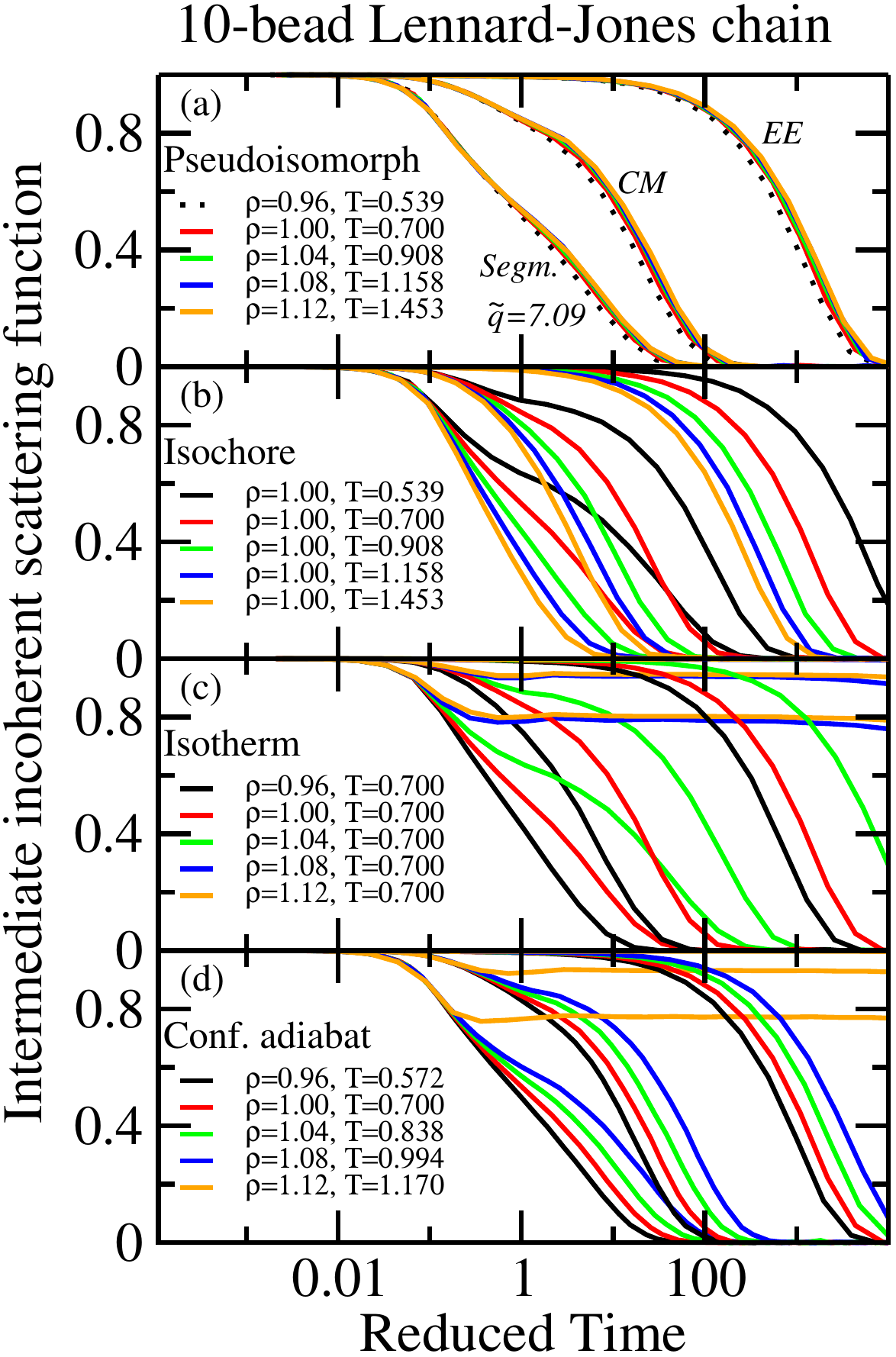}
		\caption{\label{fig:Fs}
			Incoherent intermediate scattering functions of the two models evaluated along 
			(a) the pseudoisomorphs generated as described in the text,
			(b) isochores with the same range of temperatures as in (a), 
			(c) isotherms with the same range of densities as in (a), 
			(d) constant $\Sex$ lines, i.e., configurational adiabats, in which the excess entropy is defined by subtracting the {\it atomic} ideal gas entropy. Only along the pseudoisomorphs is there collapse of the dynamics to a good approximation. 
			For the 10-bead LJC model the data in (a) for the lowest density is plotted with dotted lines to indicate that the virial is negative here ($-0.1$ in LJ-units), which for systems with isomorphs is where deviations from the predicted invariants often become significant \cite{IV} (Segm. are the segmental, CM the center-of-mass, and EE the end-to-end vector intermediate scattering functions). $\tilde q$ are the the reduced wave vectors.  
                      }
	\end{figure*}

In order to investigate whether the dynamics is invariant along the pseudoisomorphs identified as described above we calculated the incoherent intermediate scattering function corresponding to the first maximum of the structure factor (Figs. \ref{fig:Fs}(a)). For comparison, Fig. \ref{fig:Fs} gives for each model results also for: (b) an isochore, (c) an isotherm, and (d) a configurational adiabat defined as states of constant ``atomic'' $\Sex$, i.e., for which the subtracted entropy is that of the ideal gas of the \emph{atoms} constituting the molecules. Only along the pseudoisomorphs is there approximate identity of the reduced-unit dynamics. 

Is the quenching to inherent states needed in the procedure described above or would it suffice to scale the instantaneous normal modes? We investigated this for the smallest density changes for the two molecular models and found the predicted temperatures to be, respectively, 4.6\% (ASD) and 9.3\% (LJC) too small. We do not fully understand this finding, which indicates that the separation between scaling and non-scaling modes is not complete for the instantaneous normal modes. Supporting this, we find that for the Kob-Andersen system for which there are only scaling modes, one can in fact use the instantaneous normal modes to identify the isomorphs; in this case even scaling of the reduced forces may be used to identify isomorphs (unpublished). -- In connection with alternative methods for identifying lines of invariant dynamics we would like to draw attention to the recent paper by Wang and Xu, which in a somewhat different context argued that the density dependence of the glass transition temperature of LJ-type systems can be predicted from properties of the zero-temperature glasses \cite{wan14}.

In summary, a method is now available for identifying pseudoisomorphs from a single equilibrium configuration of a system of molecules with harmonic intramolecular bonds. The identity of the entire low-frequency reduced-unit spectra demonstrates that, except for the degrees of freedom associated with the springs, the potential-energy surface has the simple scaling properties assumed in the isomorph theory \cite{IV,dyr14,sch14}. This explains the existence of lines of invariant dynamics and, in particular, why excess-entropy scaling, density scaling, and isochronal superposition apply for some models that do not have strong virial potential-energy correlations.

Is it possible to generalize the isomorph theory to describe pseudoisomorphs? To generalize the isomorph-theory's defining scaling condition \cite{sch14}, $U(\bRa)<U(\bRb) \Rightarrow U(\lambda\bRa)<U(\lambda\bRb)$, imagine that configurations have degrees of freedom, some of which  scale upon a density change and some of which do not. For the harmonic-bond ASD model the former could be, e.g., the center-of-mass vector and the molecule direction vector and the latter could be the bond length, but in general there might be more than one way to define the separation. In any case, assuming that a decomposition is somehow possible of the form $\bR=(\bX,\bY)$ in which $\bX$ are the non-scaling ``spring'' degrees of freedom and $\bY$ the scaling degrees of freedom, one may expect that pseudoisomorphs exist if the following is obeyed to a good approximation: $U(\bXa,\bYa)<U(\bXb,\bYb) \Rightarrow U(\bXa,\lambda\bYa)<U(\bXb,\lambda\bYb)$.

More work remains to be done to develop a full theory of systems with pseudoisomorphs. At this point one can just conclude that lines of invariant dynamics exist for some systems without isomorphs and that these pseudoisomorphs may be identified from a single configuration. An interesting question is how far this goes: Can the method demonstrated in this paper be used more generally for identifying a relevant coarse-graining leading to pseudoisomorphs in the phase diagram? If yes, how does this connect to the current approaches to coarse graining for which important progress has recently been made \cite{bri11,cha11,noi13,dun16,lin16,she16}?

\end{document}